\documentclass[twocolumn]{aastex63}  

\usepackage[T1]{fontenc} 
\usepackage{courier}
\usepackage{amsmath}
\usepackage{enumerate}

\usepackage{dutchcal} 

\begin{document}

\shorttitle{Decaying Dark Matter and Faint X-ray Sources in XRISM}
\shortauthors{Y. Zhou, V. Takhistov, K. Mitsuda}

\title{Unlocking Discovery Potential for Decaying Dark Matter and Faint X-ray Sources with XRISM} 

\author[0000-0002-5793-554X]{Yu Zhou}
\affil{International Center for Quantum-field Measurement Systems for Studies of the Universe and Particles (QUP, WPI),
High Energy Accelerator Research Organization (KEK), Oho 1-1, Tsukuba, Ibaraki 305-0801, Japan}

\author{Volodymyr Takhistov}
\affil{International Center for Quantum-field Measurement Systems for Studies of the Universe and Particles (QUP, WPI),
High Energy Accelerator Research Organization (KEK), Oho 1-1, Tsukuba, Ibaraki 305-0801, Japan}
\affil{Theory Center, Institute of Particle and Nuclear Studies (IPNS),  High Energy Accelerator Research Organization (KEK), Tsukuba 305-0801, Japan}
\affil{Graduate University for Advanced Studies (SOKENDAI), 
1-1 Oho, Tsukuba, Ibaraki 305-0801, Japan}
\affil{Kavli Institute for the Physics and Mathematics of the Universe (WPI),   The University of Tokyo Institutes for Advanced Study, The University of Tokyo,  Kashiwa, Chiba 277-8583, Japan}

\author{Kazuhisa Mitsuda}
\affil{International Center for Quantum-field Measurement Systems for Studies of the Universe and Particles (QUP, WPI),
High Energy Accelerator Research Organization (KEK), Oho 1-1, Tsukuba, Ibaraki 305-0801, Japan}
\affil{National Astronomical Observatory of Japan, 2-21-1 Osawa, Mitaka, Tokyo 181-8588, Japan}

\email{zhouyu@post.kek.jp} 
\email{vtakhist@post.kek.jp}
\email{kmitsuda@post.kek.jp}

\begin{abstract}
Astrophysical emission lines arising from particle decays can offer unique insights into the nature of dark matter (DM). Using dedicated simulations with background and foreground modeling, we comprehensively demonstrate that the recently launched XRISM space telescope with powerful X-ray spectroscopy capabilities is particularly well-suited to probe decaying DM, such as sterile neutrinos and axion-like particles, in the mass range of few to tens of keV. We analyze and map XRISM's DM discovery potential parameter space by considering Milky Way Galactic DM halo, including establishing an optimal line-of-sight search, as well as dwarf galaxies where we identify Segue 1 as a remarkably promising target. We demonstrate that with only 100 ks exposure XRISM/Resolve instrument is capable of probing the underexplored DM parameter window around few keV and testing DM couplings with sensitivity that exceeds by two orders existing Segue 1 limits. Further, we demonstrate that XRISM/Xtend instrument sensitivity enables discovery of the nature of faint astrophysical X-ray sources, especially in Segue 1, which could shed light on star-formation history. We discuss implications for decaying DM searches with improved detector energy resolution in future experiments.\end{abstract}


\section{INTRODUCTION}
\label{section:1}

Dark matter (DM) comprises about $\sim85\%$ of all the matter in the Universe (see
e.g.~\citep{Bertone:2004pz} for review). However, all the knowledge of DM thus far originates only from its gravitational interactions.
Unraveling the mysterious nature of DM remains among the most pressing  open problems
in science.

While DM particles are expected to be long-lived, in multitude of theories DM can decay with distinct powerful signatures well-suited for indirect DM detection. Especially promising are DM decay processes resulting in monoenergetic photons that can be sensitively distinguished from typical smooth astrophysical backgrounds. Among well motivated decaying DM candidates are sterile neutrinos and axion-like particles (ALPs). 

Sterile (or right handed) neutrinos $\nu_s$ have been intimately linked to various puzzles, such as the origin~\citep{Minkowski:1977sc,Yanagida:1979as,Gell-Mann:1979vob} of small observed neutrino masses~\citep{Super-Kamiokande:1998kpq}. We consider sterile neutrinos of mass $m_s$ to mix with active neutrinos $\nu_a$ of Standard Model with mixing angle $\sin \theta$. Sterile neutrinos constitute a prime warm DM candidate for typical masses around keV (see e.g.~\citep{Boyarsky:2018tvu}) and their
radiative decays $\nu_s \rightarrow \nu_a + \gamma$ can be efficiently detected through monochromatic
X-ray line emission originating from various astrophysical targets. A possible signal has been claimed at 3.5 keV~\citep{Bulbul:2014sua,Boyarsky:2014jta} from galaxy cluster observations that could be associated with decays of sterile neutrino DM with $m_s = 7.1$~keV and $\sin^2 2\theta = 5 \times 10^{-11}$, but this has been challenged (e.g.~\citep{Dessert:2018qih}). Other parameter space of decaying sterile neutrinos is also of interest, such as in connections with tensions in measurements of the Hubble parameter (e.g. \citep{Gelmini:2019deq,Gelmini:2020ekg}). We consider natural units $c = \hbar = 1$ throughout.

Sensitive searches of decaying keV-mass DM have been carried out using X-ray observations including Milky Way's Galactic halo~(e.g.~\citep{Sekiya2016,Perez:2016tcq,Roach:2022lgo,Dessert:2023vyl,Krivonos2024}), Perseus galaxy cluster~(e.g.~\citep{Tamura:2014mta,Hitomi:2016mun}), Bullet cluster~(e.g.~\citep{Riemer-Sorensen:2015kqa}) and Local Group with Andromeda (M31) galaxy~(e.g.~\citep{Horiuchi:2013noa,Ng:2019gch})
can provide stringent tests of DM production mechanisms.
Sterile neutrino DM produced
via non-resonant active-sterile oscillations via Dodelson-Windrow (DW) mechanism~\citep{Dodelson:1993je} is already strongly constrained as the dominant DM component.
However, allowed sterile neutrino DM parameter space can be significantly modified depending on production, such as when significant lepton asymmetry~\citep{Shi:1998km} or additional neutrino self-interactions~(e.g.~\citep{DeGouvea:2019wpf,Chichiri:2021wvw,Bringmann:2022aim}) are present.
Recently, intriguing novel X-ray and gravitational wave coincidence signatures have been put forth for decaying sterile neutrino DM originating from evaporating early Universe black holes independently of couplings~\citep{Chen:2023lnj,Chen:2023tzd}. We note that serile neutrinos could themselves serve as excellent probes of early cosmological epochs~\citep{Gelmini:2019esj,Gelmini:2019wfp,Gelmini:2019clw,Gelmini:2020duq,Chichiri:2021wvw}.

ALPs constitute another motivated decaying keV-scale DM candidate (e.g.~\citep{Jaeckel:2014qea,Higaki:2014zua}).
A prominent feature of pseudoscalar ALPs is their coupling to photons $(1/4)g_{a\gamma\gamma}F^{\mu\nu}\tilde{F}_{\mu\nu}$. Decays to photons of ALP DM $a \rightarrow 2\gamma$ have been associated with the claimed putative 3.5 keV signal for ALP mass $m_a = 7.1$ keV and coupling $g_{a\gamma\gamma} \sim \textrm{few} \times 10^{-18}$~GeV$^{-1}$ (e.g.~\citep{Jaeckel:2014qea,Higaki:2014zua}). Such ALPs constituting cold DM can be produced via misalignment mechanism in the early Universe~(e.g.~\citep{Arias:2012az}). 
For strongly coupled keV-scale ALPs, even subdominant irreducible DM density contributions can be detected in X-rays~\citep{Langhoff:2022bij}. 
Prominent X-ray signatures can also arise from relativistic ALPs produced in various scenarios from transient sources contributing to diffuse axion background~\citep{Eby:2024mhd}.

In this work we comprehensively analyze discovery potential for keV-scale decaying DM with the Resolve instrument onboard X-Ray Imaging and Spectroscopy Mission (XRISM)~\citep{Tashiro2018,XRISMScienceTeam:2020rvx} that has been successfully launched on September 7, 2023 and is a successor of ASTRO-H (Hitomi) mission  that operated in 2016 and prematurely concluded observations. 
The high-energy resolution X-ray spectroscopy with Resolve using an X-ray microcalorimeter array (full width-half maximum, FWHM $\sim$ 5 eV) offers unique opportunities to probe astrophysical source emission as well as decaying DM, while Xtend, the wide-field CCD-resolution spectroscopy (FWHM $\sim$ 200 eV), provides monitor observations of the sky wider than Resolve.
Early preliminary estimates of XRISM/Resolve sensitivity have been discussed for dwarf galaxies (DG) in~\citep{Ando:2021fhj} and Galactic halo\footnote{In Ref.~\citep{Dessert:2023vyl} XRISM projections were obtained using
observed background rates from Hitomi and considering only open gate valve (GV).}
~\citep{Dessert:2023vyl}. Our analysis significantly expands on and improves multiple key aspects to chart discovery potential for XRISM, including by 
using dedicated simulations, spectra analyses with background and foreground modelling as well as consideration of multiple distinct targets to optimize the search for decaying DM. More so, we discuss opportunities for XRISM to detect faint astrophysical X-ray sources challenging to test otherwise.

\section{Decaying DM}
\label{section:2}

For keV-scale sterile neutrino DM dominant decays are $\nu_s \rightarrow \nu_a + \gamma$, resulting in monochromatic X-ray photons with energies $E_{\gamma} = m_s /2$. The channel decay rate is~\citep{Shrock:1974nd,Pal:1981rm}
\begin{eqnarray}
\centering
\Gamma_{\nu_s \rightarrow \gamma \nu_a} = 1.38\times 10^{-32} \Big(\frac{\sin^2 2\theta}{10^{-10}}\Big) \Big(\frac{m_s}{1~\textrm{keV}}\Big)^5~\textrm{s}^{-1}~,
\end{eqnarray}
considering Majorana sterile neutrinos.
For keV-scale pseudoscalar ALPs $m_a \ll m_e$ compared to electron mass $m_e$ and the decays proceed via $a \rightarrow 2\gamma$. Neglecting loop contributions from ALP-electron couplings, the rate is
\begin{eqnarray}
\centering
\Gamma_{a \rightarrow \gamma\gamma} = 7.56\times 10^{-31} \Big(\frac{g_{a\gamma\gamma}}{10^{-17}~\textrm{GeV}^{-1}}\Big)^2 \Big(\frac{m_a}{1~\textrm{keV}}\Big)^3~\textrm{s}^{-1}~,
\end{eqnarray}

The DM lifetime $\tau = 1/\Gamma$ can be directly compared to the age of the Universe.
Note that decaying DM lifetime interpretation differs by a factor of two between sterile neutrinos and ALPs due to additional photon emission.

\subsection{Galactic halo}
\label{section:2.2.1}

DM accumulates around primordial overdense regions to form halo and subhalo structures under gravitational collapse. The DM halo of the Milky Way provides a unique cosmic laboratory to probe the particle nature of DM. Assuming the collisionless DM species are in gravitational equilibrium with the gravitational potential of the Galactic DM halo, even though the photons produced by DM decays in our scenario are monochromatic, the resulting line signals are broadened due to Doppler broadening from the DM velocity dispersion. 

\begin{figure*}
    \centering
    \includegraphics[width=1.6\columnwidth]{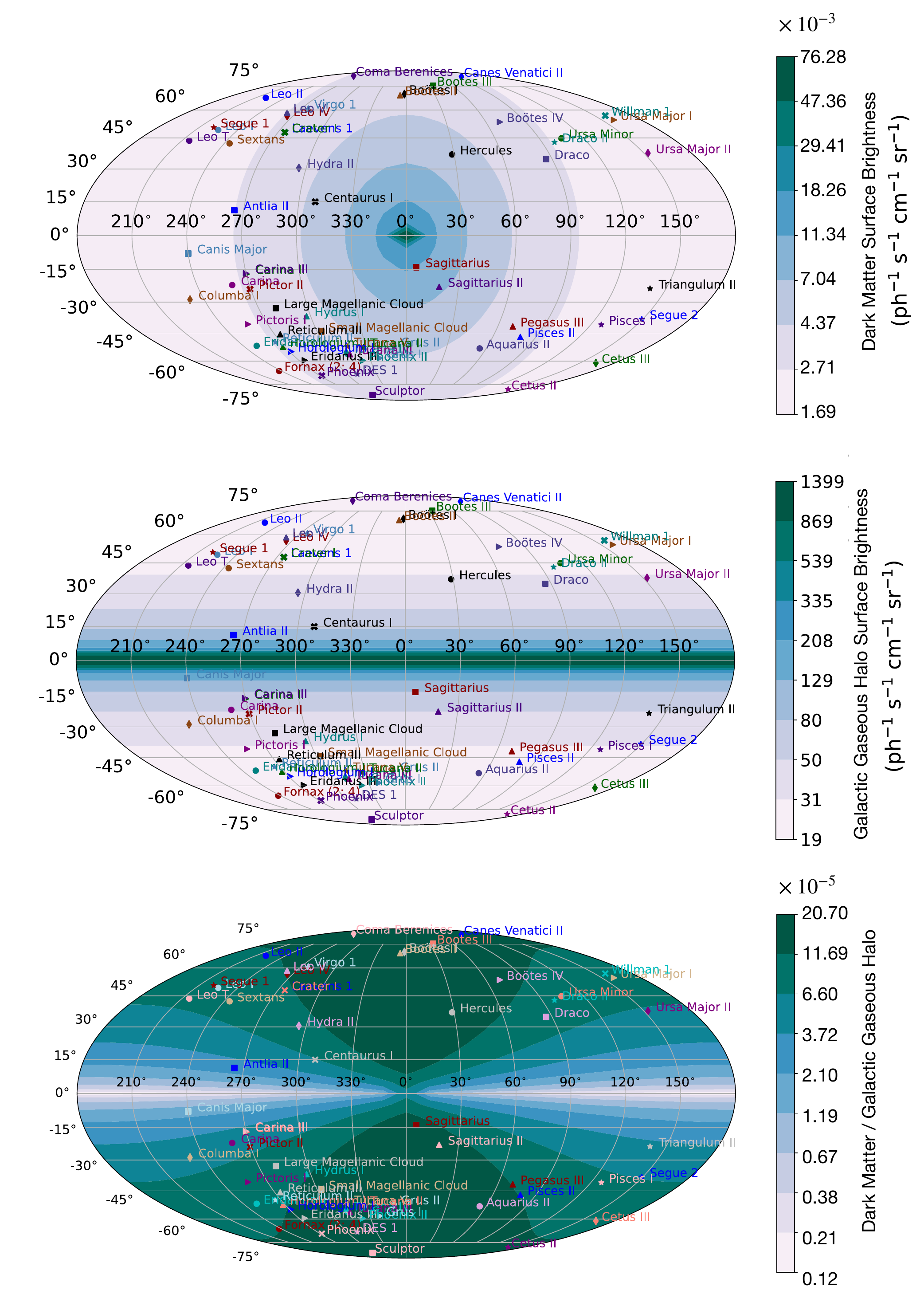}
    \caption{All-sky emission maps plotted in Galactic coordinates, along with reference positions of DGs. [Top] The DM surface brightness distribution assuming that sterile neutrinos constitute all of the DM abundance in the Milky Way Galactic DM halo. [Center] The Galactic hot gaseous halo distribution assuming disk profile. [Bottom] Map of the ratio of DM to hot gas surface brightness. Note that in center and bottom figures only X-ray emission from the halo hot gas is displayed. Selection of optimal observation directions requires proper consideration of X-ray emission from various bright sources, such as the North Polar Spur (e.g.  \citep{Snowden1997}).}
    \label{fig:Galactic LOS dependence}
\end{figure*}

However, the Galactic hot gaseous halo contributes a photon background whose spatial distribution differs from that of the DM. Therefore, there exists optimal line-of-sights where the contribution of DM signals to photon background is maximized. In our analysis, we compute the ratio of the two halo components and illustrate the optimized region for DM detection.

We model DM density distribution of the Galactic DM halo using Navarro-Frenk-White (NFW) profile~\citep{Navarro:1995iw}
\begin{eqnarray}  \label{eq:dmprof}
\centering
\rho_{\rm DM}(r) = \dfrac{\rho_s}{(r / r_s)  (1+ r/r_s)^2}~,
\end{eqnarray}
with $\rho_s = 6.6\times 10^6\rm\ M_{\odot} / kpc^3$ and $r_{\rm s} = 19.1\rm\ kpc$.
We can also consider a more general form of the DM profile, $\rho_{\rm DM}(r) = \rho_0 (r/r_s)^{-\gamma} (1+(r/r_s)^{\alpha} )^{(\gamma - \beta)/\alpha} $, where $(\alpha, \beta, \gamma) = (1,3,1/2)$ represents a more weakly-cusped DM halo, and $(\alpha, \beta, \gamma) = (2,5,0)$ represents a more cored DM halo (Plummer model). We have verified that the Doppler broadening of signal line varies no larger than $\mathcal{O}(1)$\% for different choices of considered DM profiles. We have also confirmed that the impact of DM profile on the final sensitivity forecast results is not significant.

Finite velocity dispersion of the Milky Way DM in the Galactic frame will result in Doppler broadening 
\begin{eqnarray}
\label{eq:dopper_broadening_halo}
\centering
\mathcal{f}(E, r) = \frac{4}{m_s} \frac{\int_0^{\infty} ds \rho_{\rm DM}(r) f(v(E), r)  }{\int_0^{\infty} ds \rho_{\rm DM}(r)}  ,
\end{eqnarray}
where the $f(v(E), r)$ is the DM velocity distribution projected along the line-of-sight under the assumption of a homogeneous and isotropic DM velocity distribution for a collisionless DM species in gravitational equilibrium with a gravitational potential~\citep{Dehnen2006} 
\begin{equation}  \label{eq:vel}
f(v, r) = \dfrac{e^{- v^2 / v_0(r)^2 }}{\sqrt{\pi} v_0(r)}~,
\end{equation}
where $v_0(r)^2 = 2 V_c^2(r) /(\gamma - 2 \alpha)  $,  $V_c^2(r) = G M_{\rm tot}(r)/r$ with $G$ being the gravitational constant. Here, $V_c$ is the circular velocity as a function of the radius $r$, with $M_{\rm tot} (r)$ being the mass enclosed within radius $r$. The coefficients $\alpha$ and $\gamma$ are defined by $\alpha = r \partial_r V_c (r)/V_c (r), \gamma = - r \partial_r \rho_{\rm DM} (r)/\rho_{\rm DM} (r)$, with $\rho_{\rm DM} (r)$ the DM density profile as a function of distance from the galaxy center.
We implement velocity distribution by solving Eq.~\eqref{eq:vel} assuming the DM distribution of Eq.~\eqref{eq:dmprof} for mass enclosed at radius $r$.
We have verified instead that an approximately constant velocity distribution of $v_0 = 220$ km/s everywhere in Galaxy following circular velocity curves~\citep{Eilers:2019} does not significantly affect our results.

The Doppler broadening of the signal line due to the Galactic DM halo depends on line-of-sight and approximately scales with the energy $\sigma \sim 5 \times 10^{-4} E$. For DM mass around 7 keV, the FWHM of the signal line driven by DM halo Doppler broadening is found to be around 4 eV.

Observations have suggested that the Galactic hot gaseous halo has a vertical exponential dependence on the distance away from the Galactic plane, with a temperature scale height of $\sim 1.4$ kpc and a density scale height of $\sim 2.8$ kpc, and the temperature and density at the Galactic plane $T_0 = 3.6\times 10^6$ K and $\rho_0 = 1.4\times 10^{-3}\rm\ cm^{-3}$ \citep{Yao2009}. We numerically compute the photon surface brightness map according to the Galactic disk model for each line-of-sight and integrate the emission in the Galactic halo up to 100 kpc. The emissivity of the halo plasma is computed using \texttt{APEC} code assuming solar abundance of the gas \citep{Anders&Grevesse1989}. 
Assuming sterile neutrino mass $m_a = 7.1$ keV and sterile-active mixing angle $\sin^2(2\theta) = 10^{-6}$, we create a forecast map of the DM emission surface brightness map and calculate the DM-to-gas surface brightness ratio for all line-of-sights, as shown in Figure \ref{fig:Galactic LOS dependence}.
 
\subsection{Dwarf galaxies}
\label{section:2.2.3}

DGs are the most DM-dominated systems with low quantities of photon emission, which marks them as pristine targets for exploring indirect DM signatures like decays. 
Further enhancement of DM signals can be expected if appropriate line-of-sight is selected where both DG and Galactic DM halo contributions are combined. Given that the mass-to-light $M/L$ ratio and spatial extension varies among galaxies, particular DGs constitute preferred targets for specific energy bands and observing telescopes. As we discuss, among various DGs Segue 1 is an optimal target for decaying DM search by XRISM.

\begin{figure}
    \centering
    \includegraphics[width=1\columnwidth]{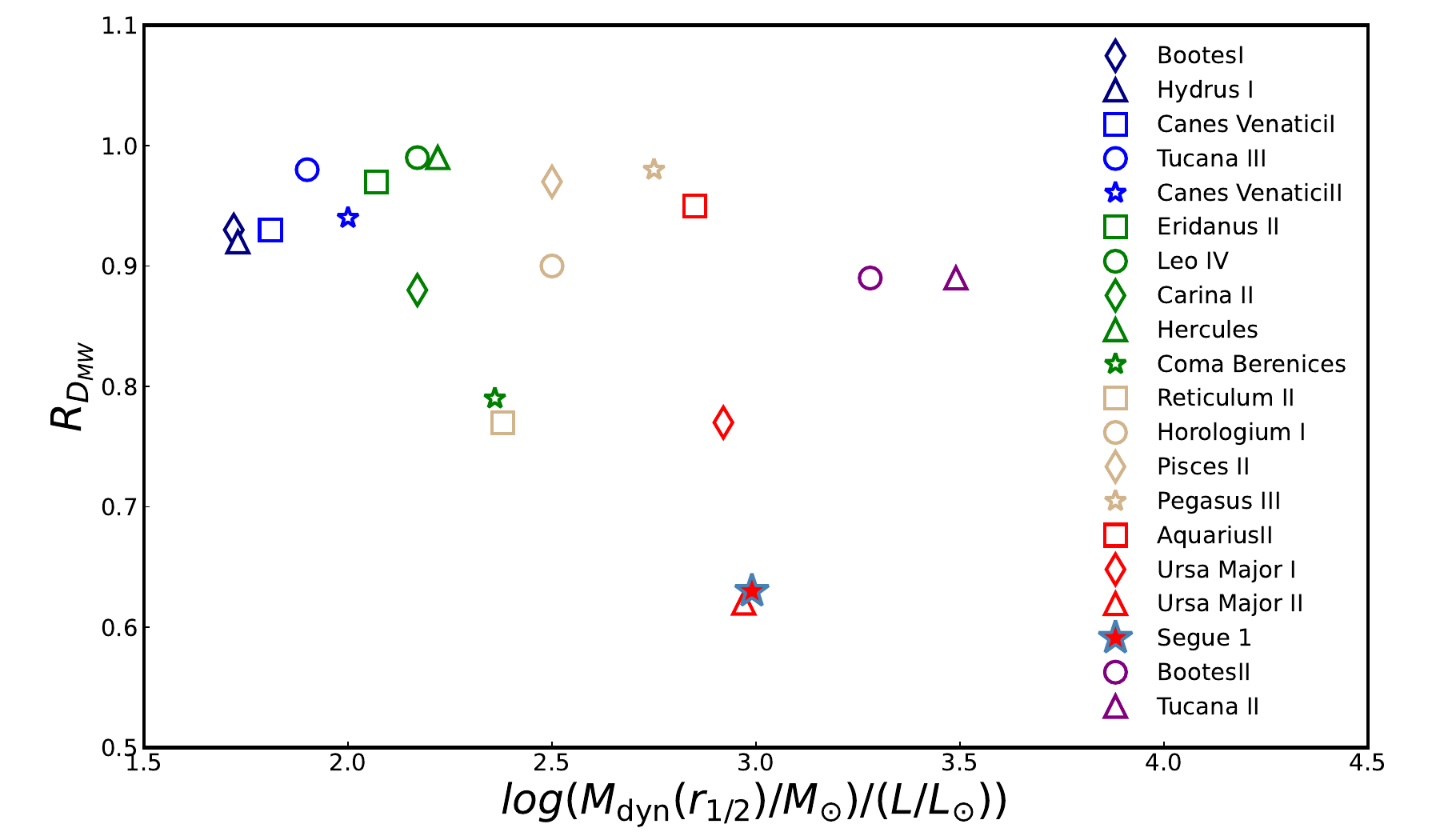}
    \caption{Distribution of 
the $D$-factor ratio $R_{D_{\rm MW}}$ of the Milky Way DM halo $D$-factor $D_{\rm MW}$ along the line-of-sight to the sum of $D$-factor including $D_{\rm DG}$ of the DG itself as a function of the dynamical mass-to-light ratio within the half-light radius $(M/L)$ in the V-band and normalized to the solar mass $M_{\odot}$ and solar luminosity $L_{\odot}$.
The DGs with higher $M/L$ ratio and lower $R_{D_{\rm MW}}$ constitute particularly favorable targets for decaying DM search.}
    \label{fig: D-factor vs. M-L ratio}
\end{figure}

To identify favorable DG targets for decaying DM search, we consider mass-to-light and $D$-factor ratios as summarized in Fig.~\ref{fig: D-factor vs. M-L ratio}. The mass-to-light ratio $M/L$ within the 3D half-light radius $r_{1/2}$ and the total V-band luminosity $L$ have been analyzed for various DGs~\citep{Battaglia2022, Ji2021, Torrealba2016, Munoz2018, Torrealba2016, Koposov2018, Cicuendez2018, Higgs2021, Collins2020, Collins2021, Cook1999, Crnojevic2014, Geha2010, Geha2006, McConnachie2012}.
Considering the dynamical mass measured from stellar motion $M_{\rm dyn}$, one can obtain dynamical mass-to-light ratio within the half-light radius $M/L \equiv M_{\rm dyn} (r_{\rm 1/2}) / (L/2)$.
Majority of the DGs are found to have $M/L\gtrsim 10$, with $M/L$ tending to increase at lower $L$. Therefore, lower luminosity DGs are more dominated by the DM in mass. 

The $D$-factor accounts for the distribution of DM in an astrophysical system to determine the strength of the emission signal from decaying DM
\begin{eqnarray}
\centering
D_{\rm DG} = \int_{\Delta \Omega} d\Omega \int dl \rho_{\rm DM}(r(l, \Omega)),
\end{eqnarray}
where $\Omega$ is the solid angle of the chosen field of view and $l$ is the distance along the line-of-sight.
The $D$-factor ratio of the Milky Way to a given DG can be found from 
\begin{equation}
    R_{D_{\rm MW}} = \dfrac{D_{\rm MW}}{D_{\rm MW} + D_{\rm DG}}~.
\end{equation} 
This characterizes how the DG DM emission is enhanced compared to the DM signal stemming from the Galactic DM halo~(see e.g.~\citep{Ando2021}).
Smaller $R_{D_{\rm MW}}$ indicates more significant DG DM signal enhancement compared to the Milky Way.

In Fig.~\ref{fig: D-factor vs. M-L ratio} we display $D$-factor ratio versus $M/L$ ratio. The DGs with the highest $R_{D_{\rm MW}}$ values are Segue 1 and Ursa Major II, having $R_{D_{\rm MW}}=$ 0.56 and 0.54, respectively.
DGs with the highest $M/L$ ratios and smallest $R_{D_{\rm MW}}$ values constitute particularly favorable targets for decaying DM search. 

Other considerations further indicate that DG Segue 1 is an especially favorable target for our analysis. Taking into account XRISM/Resolve capabilities, which can only observe a very limited field of view ($\sim2.9\times 2.9\rm\ arcmin^2$), half-light radius or DM concentration of the DG is another essential factor for optimal target selection. Segue 1 has a small estimated half-light radius of $\sim$4.3 arcmin, ideally matching with the XRISM/Resolve field of view. More so, Segue 1 DG has a favorable sky location such that DM-to-gas surface brightness ratio is relatively high but it is still far from the Galactic center on the sky map, meaning it is less affected by the photon contamination from the Galactic hot gaseous halo and gas bubble identified by eROSITA~\citep{Predehl2020,Gupta2023,Liu2024}.

\subsection{Segue 1 }
\label{section:2.2.4}

Segue 1 is the least luminous of the ultra-faint DGs discovered around the Milky Way. With a mass-to-light ratio considering V-band luminosity of $M/L_{V} = 2440_{-1775}^{+1580}$, this DG is found to be DM-dominated at a high significance~\citep{Geha2009}. 
The mean heliocentric recession velocity of $206\pm 1.3\rm\ km\ s^{-1}$ is measured from 24 stars identified as members of the DG and the internal velocity dispersion is measured for $4.3\pm 1.2\rm\ km\ s^{-1}$~\citep{Geha2009}.  The DM distribution of the Segue 1 DG is calculated based on the spherical NFW profile.
We adopt the $D$-factor within 5 arcmin radius, $9.94\times 10^{16}\rm\ GeV/cm^2$, according to the value in Figure 6 in \citep{Evans2016}. 
Previously, Segue 1 constraints on sterile neutrino decaying DM were obtained using data from a short $\sim 5$~ks X-ray observation period by the Swift telescope~\citep{Mirabal2010}. As we demonstrate, XRISM can dramatically improve on these results.

Stellar chemical analyses also appear to indicate that Segue 1 remains as a fossil galaxy that might have only experienced a single star formation activity phase~\citep{Frebel2014}. Hence, besides DM, discovery opportunities of faint X-ray binary sources in Segue 1 that could potentially relay key information about the early epoch star formation also make it an attractive target source for deep observations with sensitive detectors. 

\begin{figure}
\centering
 \includegraphics[width=\columnwidth]{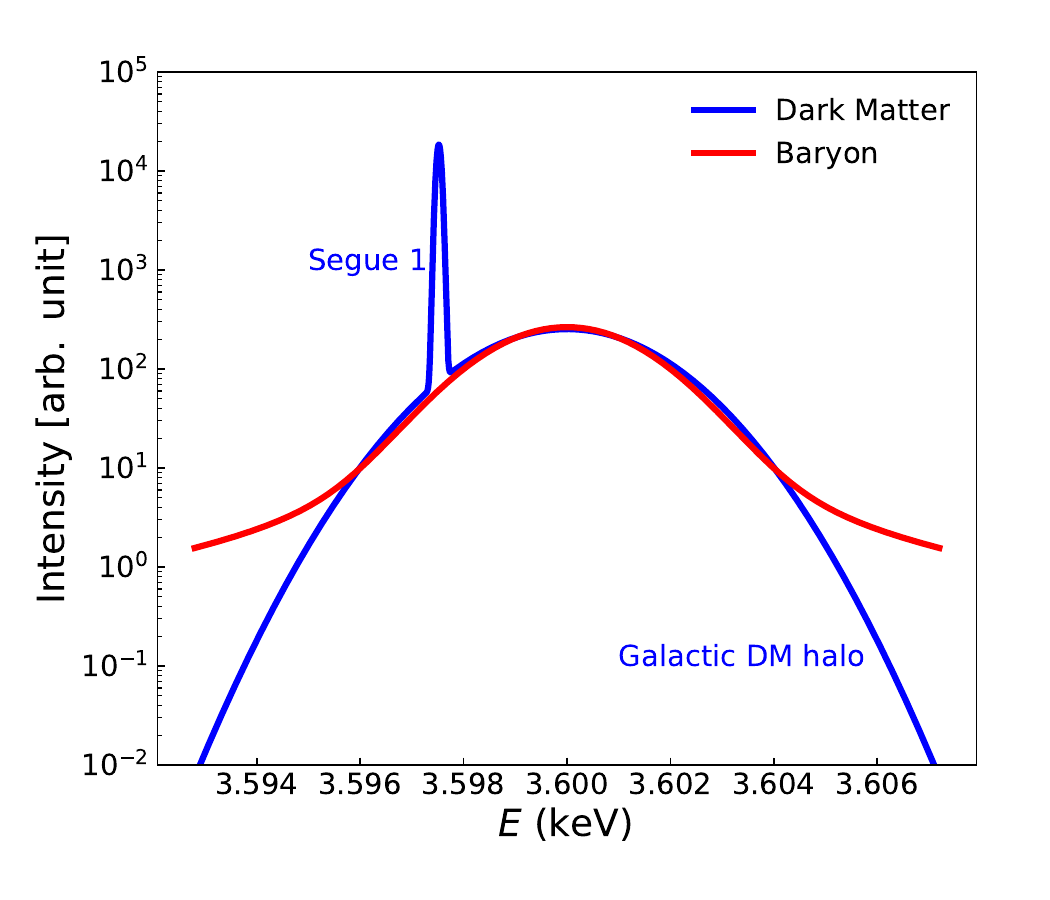}
 \caption{Comparison between  decaying DM signal and the atomic emission from Segue 1 DG along the line-of-sight assuming $m_s = 7.1$~keV, highlighting features that can be distinguished by a detector with an appropriate energy resolution. 
 (blue) Combined decaying DM signal as observed from Segue 1 DG along the line-of-sight. The narrow line originates from the Segue 1 DM halo, and the broadened line originates from the Galactic DM halo within the telescope field of view. (red) Voigt profile presenting a hypothetical atomic emission line from the Galactic hot gaseous halo. Assumed relevant parameters such as turbulence velocity, atomic number and the transition rate of atomic decays are described in Sec.~\ref{section:2.2.4}. }
 \label{fig:line profile}
\end{figure}

The emission spectrum contributions of decaying DM $\chi$ from a DG and the Galactic DM halo along the line-of-sight can be modeled as
\begin{eqnarray}
\label{eq:line_model}
\centering
\frac{d\phi}{dE} = \frac{\Gamma_{\chi}}{4\pi m_{\chi}} \Big(\frac{dN_{\rm dec}}{dE}\Big|_{\rm DG} D_{\rm DG} +  \frac{dN_{\rm dec}}{dE}\Big|_{\rm MW} D_{\rm MW}\Big).
\end{eqnarray}
Here, $dN_{\rm dec}/dE$ is DM decay energy spectrum. Without taking into account Doppler shift or broadening effects, it is just a delta function centered at monochromatic energy $E = m_{\chi}/2$. 
In Fig.~\ref{fig:line profile}
we display the line profile emission of the DM signal from Segue 1 consisting of a red-shifted line broadened according to the internal velocity dispersion of Segue 1 superimposed on top of the Galactic DM emission contribution with line width dominated by the Doppler broadening of more than $\sim4$ eV, assuming DM mass of $m_{\rm \chi} = 7.2\rm\ keV$.  

As we demonstrate in Fig.~\ref{fig:line profile}, the Segue 1 DM signal features could be well distinguished with high energy resolution and statistics observations from atomic signals whose energy distribution is subject to Voigt profile. The Voigt profile of an atomic line is determined by transition rate between the energy levels and thermal temperature of the emitter
\begin{eqnarray}
\centering
V(E, \sigma, \gamma) = \int_{-\infty}^{+\infty} G(E^{'}, \sigma) L (E-E^{'}, \gamma) dE^{'},
\end{eqnarray}
where $G(E, \sigma)$ is a Gaussian distribution of energy $E$ with a width of $\sigma = E \sqrt{2kT/m_{\rm ion} + v_{\rm turb}^2}$ characterizing the thermal broadening due to the gas motion, and $L(E, \gamma)$ is the Lorentzian profile 
$L(x, \gamma)=\gamma/(\pi (\gamma^2 + x^2))$ with $\gamma = \nu$ being the transition rate frequency. Here, for the atomic line profile shown in Fig.~\ref{fig:line profile}, 
we consider reference parameters corresponding to common properties of the hot gases in the Milky Way halo 
with the temperature of the hot gas $T = 10^7\rm\ K$ \citep{Nakashima2018}, turbulence velocity $v_{\rm turb} = 100\rm\ km/s$ \citep{Li2017}, 
ion mass $m_{\rm ion} = 39.098 m_{\rm p}$ with $m_p$ being the proton mass and a transition rate 
of $\nu = 3.43\times 10^{14}\rm\ Hz$. The choice of atomic transition parameters are referred to  
atomic database \ttfamily{ATOMDB(version 3.0.9)}\normalfont \footnote{http://www.atomdb.org/Webguide/webguide.php} 
but assigned with fictitious values. The ion mass has assumed the mass of Potassium and the transition rate is close 
but not equal to the Ar XVII (level 1$\to$7, $\nu = 1.09\times 10^{14}\rm\ Hz$).

\section{Foreground and background Emission}
\label{section:3}

In our analysis, we model background and foreground emission by accounting for the following major contributions. In Fig.~\ref{fig:Galactic LOS dependence} we display the all-sky map of Galactic gas halo.
 
--\textit{Local hot bubble (LHB) and solar-wind charge exchange (SWCX):} LHB has been discovered around the Solar System neighborhood with an irregular size of approximately $\sim 200$~pc using observations of an intense diffuse soft X-ray emission coupled with the fact that the solar neighborhood is almost completely devoid of cold gas, leading to a picture of a ``local cavity'' filled with hot gases~\citep{McCammon1990,Snowden1993,Farhang2019}. We use the collisionally-ionized diffuse gas emission spectrum (APEC) to model the LHB. We adopt the temperature corresponding to an average number 61.275 eV of the fitted \texttt{APEC} temperature to a Suzaku blank-sky X-ray background datasets \citep{Zhou2021}. The surface brightness of LHB is also set to an average value of the observed Suzaku X-ray background, 165.981 $\rm ph\ s^{-1}\ cm^{-2}\ sr^{-1}$. Since the LHB emission is considered as local around our Solar System, no Galactic absorption from the interstellar medium is applied to this component. Due to limitations of the energy resolving power of CCD detectors, the non-thermal line contributions from the SWCX process cannot be distinguished from the LHB component. Therefore, the model of LHB consisting of average parameter values as inferred from Suzaku background observations already includes the impact from SWCX. 

--\textit{Galactic gaseous halo:}
Galactic stellar formation processes can eject hot gases into surrounding environment so that hot gases accumulate near the Galactic plane to form a disk~\citep{Hagihara2010,Yao2009,Yao2012,Sakai2014}. Alternatively, shock heating driven by the DM core collapse can heat the baryons into a warm-hot phase to several million Kelvin inside the DM halo~\citep{Fang2013}. Diffuse hot gas permeating the Galactic DM halo exist in a form of an optically thin thermal plasma, of which we modeled the X-ray emission with the \texttt{APEC} model in the \textbf{XSPEC} software~\citep{xspec1996}, assuming an average temperature of $0.178\rm\ keV$ and an average surface brightness of 6.24 $\rm ph\ s^{-1}\ cm^{-2}\ sr^{-1}$ according to the earlier Suzaku X-ray background analysis by some of us~\citep{Zhou2021}. We use the \texttt{TBabs} model of \textbf{XSPEC} to account for the Galactic foreground ISM absorption, with neutral hydrogen column density assumed to the medium value considering all Suzaku background observations, i.e. $n_{\rm H} = 1.8\times 10^{20}\rm\ cm^{-2}$. 

--\textit{Extragalactic cosmic X-ray background (CXB):}
The predominant contributions to the X-ray background above 2 keV are expected to originate from the distant active galactic nuclei (AGN) and galaxies. We make use of double broken power-law to model CXB emission
\begin{equation}
\begin{cases}
      K E^{-\Gamma_1} & \text{if $E \leq E_{\rm break}$}\\
      K E_{\rm break}^{\Gamma_2-\Gamma_1}(E/\textrm{1 keV})^{-\Gamma_2} & \text{if $E > E_{\rm break}$}\\
    \end{cases}       
\end{equation}
where $K$ is the normalization and we take the photon indices fixed at $\Gamma_1 = 1.96$ and $\Gamma_2 = 1.54$ 
below the break point energy $E_{\rm break} = 1.2\rm\ keV$, and $\Gamma_{1,2} = 1.4$ above $E_{\rm break}$~\citep{Yoshino2009}. 
The normalizations of the broken power-law components are set to 3.7 and 5.7 $\rm ph\ s^{-1}\ cm^{-2}\ keV^{-1}\ sr^{-1}$ at 1 keV, 
according to the average normalization value fitted for the Suzaku X-ray background~\citep{Zhou2021}. 
We use the \texttt{TBabs} to model the Galactic foreground ISM absorption with the same neutral hydrogen 
column density as described previously.
In principle, the unresolved CXB surface brightness could vary among observations taken with different 
angular resolution and exposure time. The CXB intensity assumed for this work, which is obtained from Suzaku background, 
is in agreement with the full CXB intensity measured by Chandra COSMOS legacy survey within 5\% in 0.3-10.0 keV and 
within 12\% in 2-10 keV \citep{Cappelluti2017}. \citet{Cappelluti2017} have also compared the total unfolded CXB spectrum with other 
previous observations and illustrate the consistency among Chandra, Swift, ROSAT-ASCA, Integral, RXTE results with a 
discrepancy level no larger than 20\%.

--\textit{Faint X-ray binaries in Segue 1:}
The eROSITA all-sky survey point source catalog~\citep{Merloni2024} suggests that in the 0.5-2 keV range there is no significant detection of X-ray point source emission originating from the central region of Segue 1 DG in the field of view of XRISM/Resolve. Hence, this is a particular well-suited target for DM signal search using XRISM. However, there could exist faint X-ray sources with emission below the eROSITA all-sky survey flux threshold of $F_{0.5-2\rm\ keV} = 5\times 10^{-14}\rm\ erg\ s^{-1}\ cm^{-2}$ that are in the XRSIM's field of view. Assuming the flux of such a faint source is at the limit probed by eROSITA, we would expect $\sim$140 counts if observed with XRISM/Xtend for 100 ks exposure time. Given that the energy spectrum of the X-ray binary system is continuum, the contribution of such faint sources will still be less dominant than the total diffuse X-ray background and the non-X-ray background. Therefore, we can conclude that Segue 1 DM sensitivity reach is not expected to be significantly degraded due to the unknown faint point source contamination. 

--\textit{Non-X-ray background:}
Energetic particles, mainly protons originating from space or solar activity, can excite secondary photons inside the telescope and contribute a considerable amount of background in the observed energy spectrum \citep{Tawa2008}. We simulate non-X-ray background of the XRISM's Resolve spectrometer using dedicated calibration input \textit{resolve\_h5ev\_2019a\_rslnxb.pha}. The resulting count rate of $\sim 0.06\rm\ counts\ s^{-1}\ cm^{-2}$ is consistent with the orbit-averaged data collected by Hitomi/SXS NXB that gives $\sim 0.04\rm\ counts\ s^{-1}\ cm^{-2}$ in the energy range of 0.3-12 keV~\citep{Kilbourne2018}. 

\begin{figure}
\centering
 \includegraphics[width=1\columnwidth]{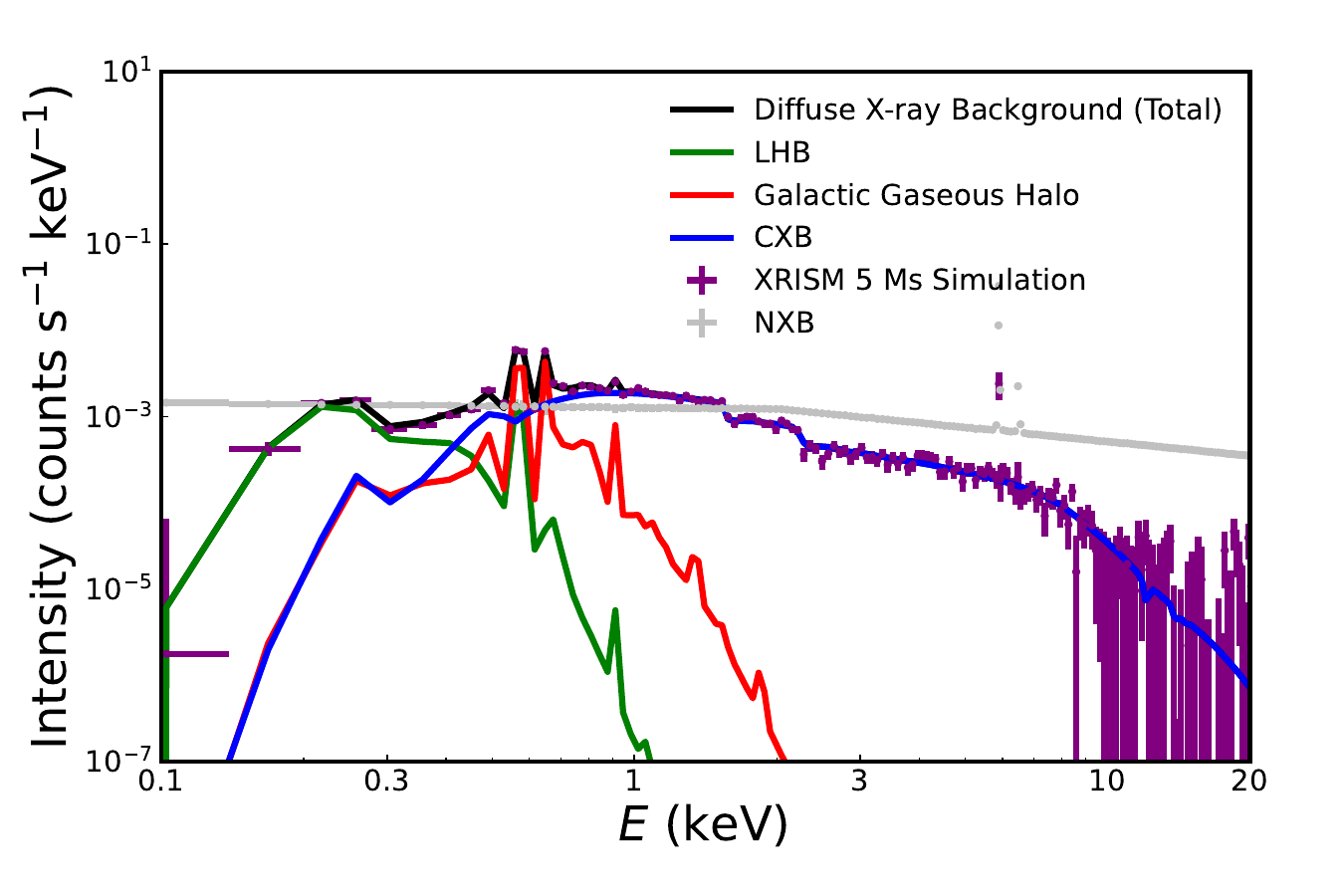}
 \caption{Simulation of  XRISM/Resolve X-ray background showcasing contributing components, considering 5 Ms exposure. Total diffuse X-ray background emission (black) including the local hot bubble (LHB), Galactic hot gaseous halo and cosmic X-ray background (CXB), local hot bubble  model used in simulation (green), galactic hot gaseous halo model used in simulation (red), cosmic X-ray background model used in simulation (blue), simulated energy spectrum of the total diffuse X-ray background using XRISM/Resolve response (purple), the non-X-ray background energy spectrum used in the simulation (gray).}
 \label{fig:background}
\end{figure}

\section{Dark matter discovery reach}

\subsection{Observation target systems}
We consider three astrophysical observation target systems and construct DM signal models for them to identify DM discovery reach sensitivity.

--\textit{Average random Galactic DM halo line-of-sights:}~The DM surface brightness is averaged for Galactic DM halo across the sky (longitude $l = 0^{\rm o} - 360^{\rm o}$ and latitude $b=0^{\rm o} - 90^{\rm o}$) in the model. We configure the temperature and the surface brightness of the Galactic foreground considering the average values obtained from the Suzaku X-ray background analysis.  

--\textit{Optimized Galactic DM halo line-of-sight:}~From the DM-to-gas ratio surface brightness map shown in Fig.~\ref{fig:Galactic LOS dependence} we have computed an optimized line-of-sight exists at $l=0^{\rm o}, b=40^{\rm o}$ where the DM-to-gas surface brightness ratio reaches maximum. However, ROSAT hot gas map at 3/4 keV \citep{Snowden1997} and eROSITA all-sky map suggest that significant amount of hot bubble (i.e. ``eROSITA bubble'') emission contaminates this region \citep{Predehl2020,Gupta2023,Liu2024}. This introduces additional complexity and can obfuscate DM signals in data analysis. Thus, by exploiting the information from both the DM-to-gas surface brightness ratio map and the observed 3/4 keV hot gas map we have identified a secondary favorable line-of-sight for DM search at $l=330^{\rm o}, b=60^{\rm o}$, where the DM-to-gas surface brightness ratio is significant and photon emission from the eROSITA bubble is low. 
We compute the DM surface brightness, Doppler broadening, and the surface brightness of Galactic hot gaseous halo by disk model specifically for this line-of-sight, and use those parameters in the X-ray background spectrum simulations as well as fit.

--\textit{Line-of-sight towards Segue 1:}~The DM signal emission originating from the line-of-sight chosen towards Segue 1 consists of two components. One is from the DM of the DG itself and the other from the Galactic DM halo along the line-of-sight, as described by Eq.~\eqref{eq:line_model}. In our analysis we have assumed that the DM associated with Segue 1 posses the same recession velocity and velocity dispersion as indicated by stars~\citep{Geha2009}. Thus, we have applied a Doppler redshift of $\sim 0.7\%$ relative to the rest-frame energy and a Doppler broadening of $\sim 0.5$ eV to the line signal from Segue 1.

\begin{figure}[t]
    \centering
    \includegraphics[width=1\columnwidth]{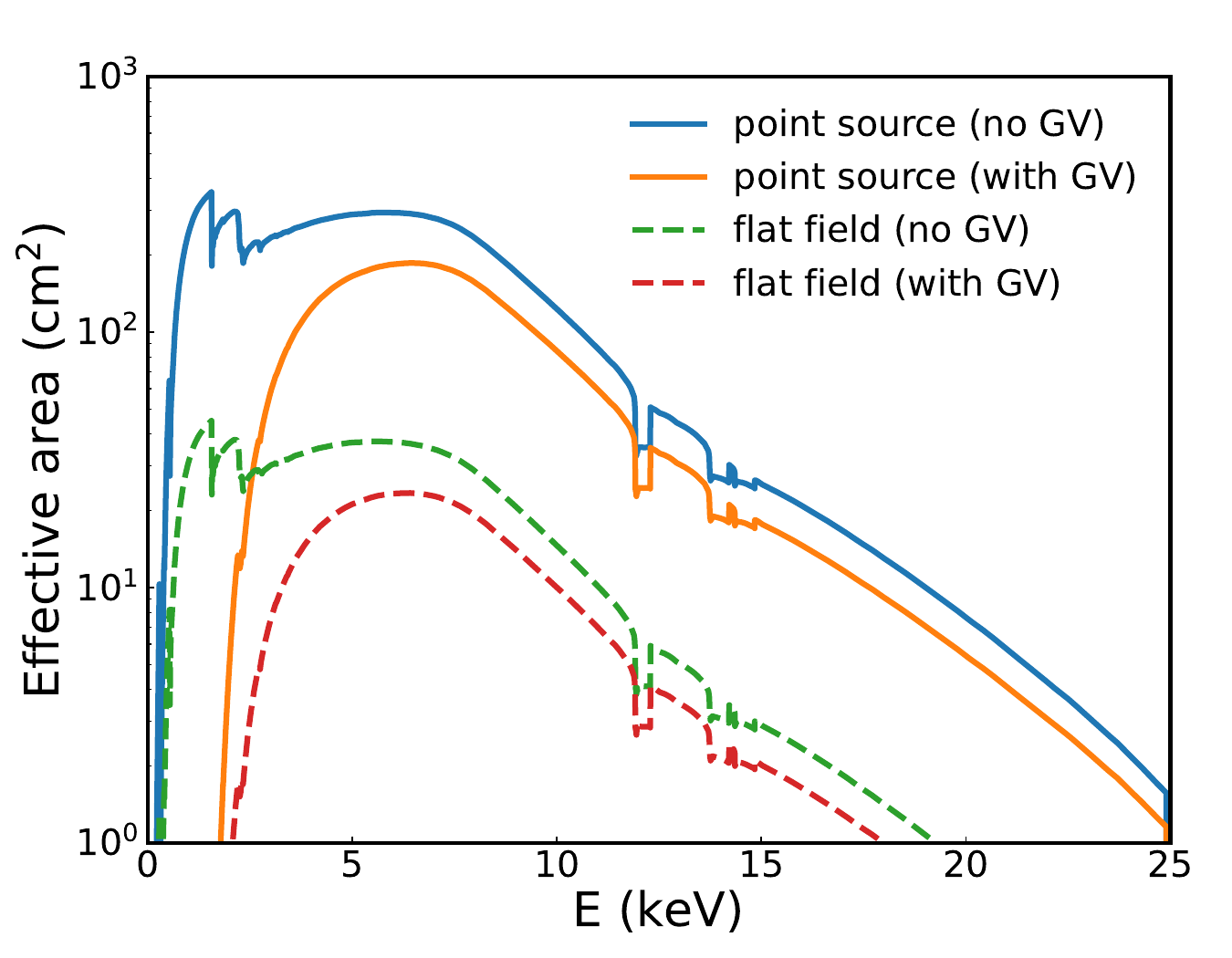}
    \caption{Effective area response of the XRISM/Resolve for point sources and flat field assuming 5 arcmin radius sky, considering the condition that gate valve (GV) is open or closed.}
    \label{fig:xrism effective area}
\end{figure}

\subsection{Simulations}

XRISM space telescope observatory is equipped with two detectors, Resolve and Xtend, adopting X-ray microcalorimeter and CCD technology, respectively. The X-ray microcalorimeter XRISM/Resolve covers a field of view of $2.9\times2.9\rm\ arcmin^2$ and an energy band 0.3-12 keV. XRISM/Resolve has an effective area of $> 210\rm\ cm^2$ at 6~keV and an absolute energy scale $<2\rm\ eV$. The unprecedented energy resolution of XRISM/Resolve facilitates very high sensitivity for DM signal detection. On the other hand, XRISM/Xtend has a field of view larger than 300 cm$^2$ at 6~keV that enables potential discovery of faint point sources. 

For spatially extended emission, we need to treat carefully the X-ray telescope's responses such as the vignetting function and the point spread functions including stray light.  In a standard X-ray spectral analysis, we construct a so-called \texttt{arf} (auxiliary response input file), which describes the effective area of the telescope as a function of X-ray energy for assumed spatial distribution of emission.  To construct \texttt{arf}, specific ray-tracing software of the telescope is required.  Instead we consider two \texttt{arfs} for extreme cases that are available, a point source at the optical axis, and a flat field extending in a 5 arc-min radius circle centered at the optical axis. In Fig.~\ref{fig:xrism effective area} we display the effective area of XRISM/Resolve \texttt{arf} inputs for a point source and a 5 arc-min radius flat field, respectively. 
For the foreground and background X-ray emission and for the DM emission from the Galactic halo,  the \texttt{arf} for 5 arc-min radius flat field is precise. However, for the DM emission from DGs, both \texttt{arfs} are not very precise. In our simulation, we adopt the flat field \texttt{arf} \textit{resolve\_flt\_spec\_no(with)GV\_20190611.arf}. In Sec.~\ref{sec:Fit_Results} we estimate the surface brightness of the DM emission corresponding to the projected upper limit of the spectral fit parameter. Assuming the spatial distribution of the DM emission expected for Segue 1 DG, we estimate that the systematic error due to the choice of the \texttt{arf} in this conversion is less than $\sim20$\%. 

We simulate the diffuse X-ray background emission for XRISM/Resolve using the 5 eV resolution response input  \textit{resolve\_h5ev\_2019a.rmf}. We consider the model \texttt{TBabs*(APEC + bknpower + bknpower) + APEC} specified in \textbf{XSPEC} and simulate the spectra with the XRISM response using \texttt{fakeit} routine. 
Here, \texttt{TBabs} models the ISM absorption in the Milky Way characterized by neutral hydrogen column density $n_{\rm H} = 1.8\times 10^{20}\rm\ cm^{-2}$, the value of which we calculate for Segue 1 line-of-sight according to HI4PI map~\citep{HI4PI2016}. The first  \texttt{APEC}  component models for the Galactic hot gaseous halo, the double broken power-law components for the CXB, and the second  \texttt{APEC}  for the local hot bubble and SWCX components. Dark matter signal model is constructed but not included in the background simulation. Parameters of these components are chosen as average values described in Sec.~\ref{section:3}.
We include the non-X-ray background in simulations by using the calibration input \textit{resolve\_h5ev\_2019a\_rslnxb.pha}.

We consider the telescope effective area for flat field with and without gate valve open, using ancillary inputs \textit{resolve\_flt\_spec\_noGV\_20190611.arf} and \textit{resolve\_flt\_spec\_withGV\_20190611.arf}, respectively. The effective area for flat field is generated with ray-tracing simulations assuming the source of emission is uniformly distributed in the 5 arcmin radius sky surrounding the telescope pointing direction. Correspondingly, the \texttt{DM} model is also computed for the 5 arcmin radius sky area to account for a reference surface brightness value. In this way, any non-uniform telescope response to the photon emitted in the off-axis regions or outside of the field of view has been taken into account in our simulations. These considerations improve the accuracy of analysis compared to a point-source response. The difference in the effective area when the gate valve is open and closed is described in detail in Ref.~\citet{Tsujimoto2018}. We simulate diffuse X-ray background for 100 ks, 5 Ms, and 100 Ms to compare the impact of exposure time. For each configuration, we generated 1000 realizations for both the photon background spectra and the non-X-ray background spectra. The non-X-ray backgrounds are subtracted later from the total diffuse background spectrum in the fitting procedure. 

We also consider how sensitivity will improve for a future experiment with 2 eV energy resolution.
For this we constructed appropriate detector response inputs. The effective area is assumed to have the same energy dependence as XRISM/Resolve, whose absolute amplitude is similar to XRISM/Resolve point source response of $\sim 300\rm\ cm^{2}$ at 6~keV, but about eight times larger than the flat field response of $\sim 40\rm\ cm^{2}$ at 6~keV.

 \begin{figure}
\centering
 \includegraphics[width=\columnwidth]{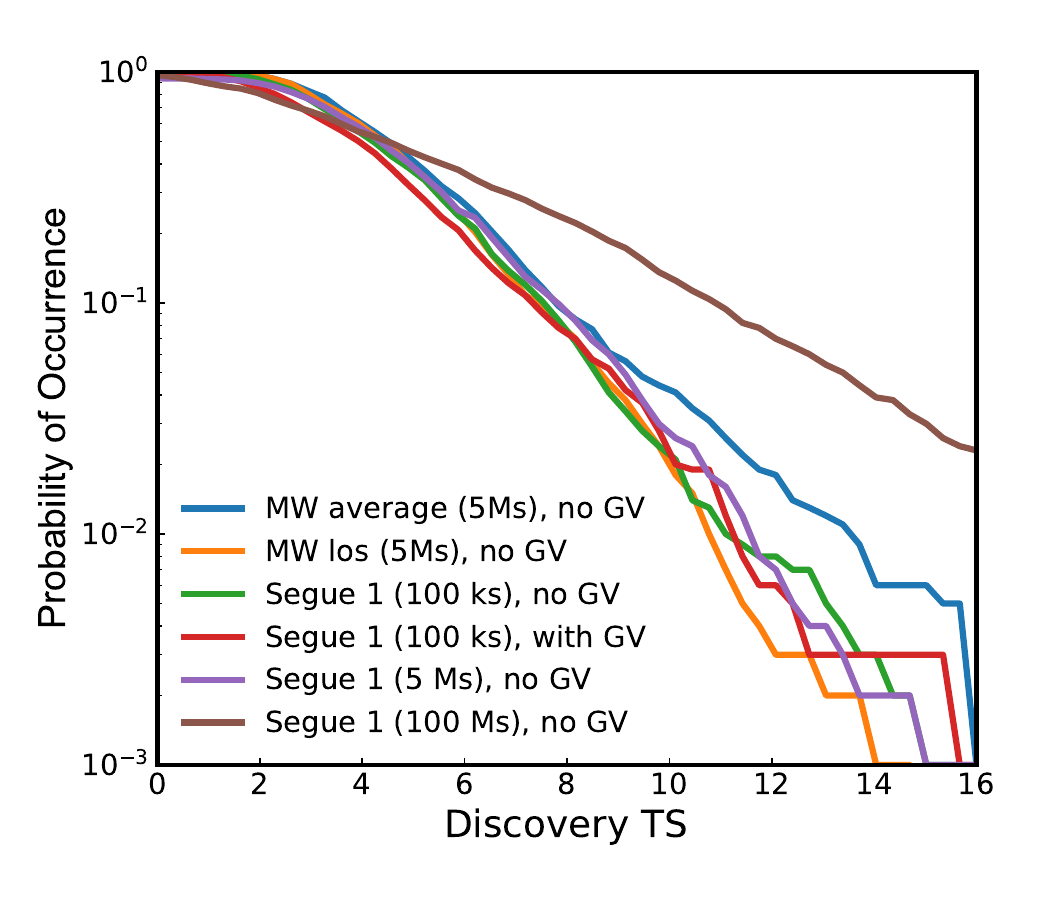}
 \caption{Probability of occurrence defined as p-values of the max($C_0 - C_{\rm min}$) distribution for 1000 simulations. The $C_0 - C_{\rm min}$ value are computed for all fits with \texttt{DM} signal model of different mass $m_s$ in the searched energy range, the maximum of which is used as the global discovery test statistic. }
 \label{fig:statistics}
\end{figure}

\subsection{Analysis fit}

We fit the simulated datasets with the model \texttt{TBabs*(APEC + bknpower + DM) + APEC} in \textbf{XSPEC}. The hydrogen column density of \texttt{TBabs} model is fixed at $n_{\rm H} = 1.8\times 10^{20}\rm\ cm^{-2}$. Temperatures of the first and second  \texttt{APEC}  components are fixed at the average numbers that were used in the simulation, while the normalizations of both components are set as free fitting parameters. We use a single broken power-law in the fit instead of the double broken power-law in simulations, because the low statistics of simulation spectra do not allow to constrain on both power-law components. Therefore, we have only kept the first broken power-law and only set the normalization as free parameter for fitting. The \texttt{DM} component is the DM model signal, which is computed independently for each DM mass $m_a$ as described in Sec.~\ref{section:2}. We fitted for the DM signal in the range of 0.3-15 keV that is within the sensitive bandwidth of XRISM/Resolve. Normalization of the \texttt{DM} component is considered as a free parameter that is fitted to find upper limit projections constrained by the background spectrum. 

In case the total photon count amount is low, we calculate the C-statistic \citep{Cash1979} instead of $\chi^2$ as 
\begin{eqnarray}
\centering
C = 2\Sigma_{i=1}^{N} (t m_i) - {S_i} \ln(t m_i) + \ln(S_i !),
\end{eqnarray}
where $S_i$ are the observed counts, $t$ is the observation time, and $m_i$ the predicted count rates based on the considered model and detector response. The C-statistic is used to maximize the likelihood for the data following Poisson distribution.
For each realization, we scan the \texttt{DM} normalization value starting from zero to a sufficiently large number and find how the C-value depends on the normalization.

Since we are searching for the signal of DM $\chi$ with mass $m_{\chi}$ over a continuous energy range, in order to assess the significance of a local deviation from the background-only hypothesis we need to also take into account the probability of such a deviation to occur anywhere within the search range, i.e. ``look elsewhere effect''. Therefore, to investigate the distribution of the global discovery test statistic (TS) we have computed the C-value curves for different DM masses $m_{\chi}$, i.e. signal energies centered around $m_{\chi} / 2$, and found the maximum value of $(C_0 - C_{\rm min})$. We define $C_{\rm min}$ as the local minimum of C-value reached in the normalization range of the scan, and $C_0$ as the C-value where normalization equals to zero. For all \texttt{DM} masses $m_{\chi}$ the values of $C_0$ are identical, since they result from the same simulation data fitted with an identical diffuse X-ray background model. Thus, we use max($C_0 - C_{\rm min}$) to characterize the global discovery TS. 

For all 1000 simulation datasets we have performed the above analysis and obtained the distribution of the global discovery TS. The probability of occurrence of max($C_0 - C_{\rm min}$) values, which is defined as the $p$-value of the max($C_0 - C_{\rm min}$) distribution, is shown in Fig.~\ref{fig:statistics}. From the probability of occurrence curve P(max($C_0 - C_{\rm min}$)), we find the max($C_0 - C_{\rm min}$) value where P=2.5\%, noted as $C_{\Delta}$. In order to determine the global 95\% confidence level sensitivity upper limit reach of \texttt{DM} normalization at each $m_{\chi}$ we search for the minimum normalization value where $(C_{\rm min} + C_{\Delta})$ is reached and compute the average for all simulations. 

\begin{figure}
    \centering
    \includegraphics[width=1\columnwidth]{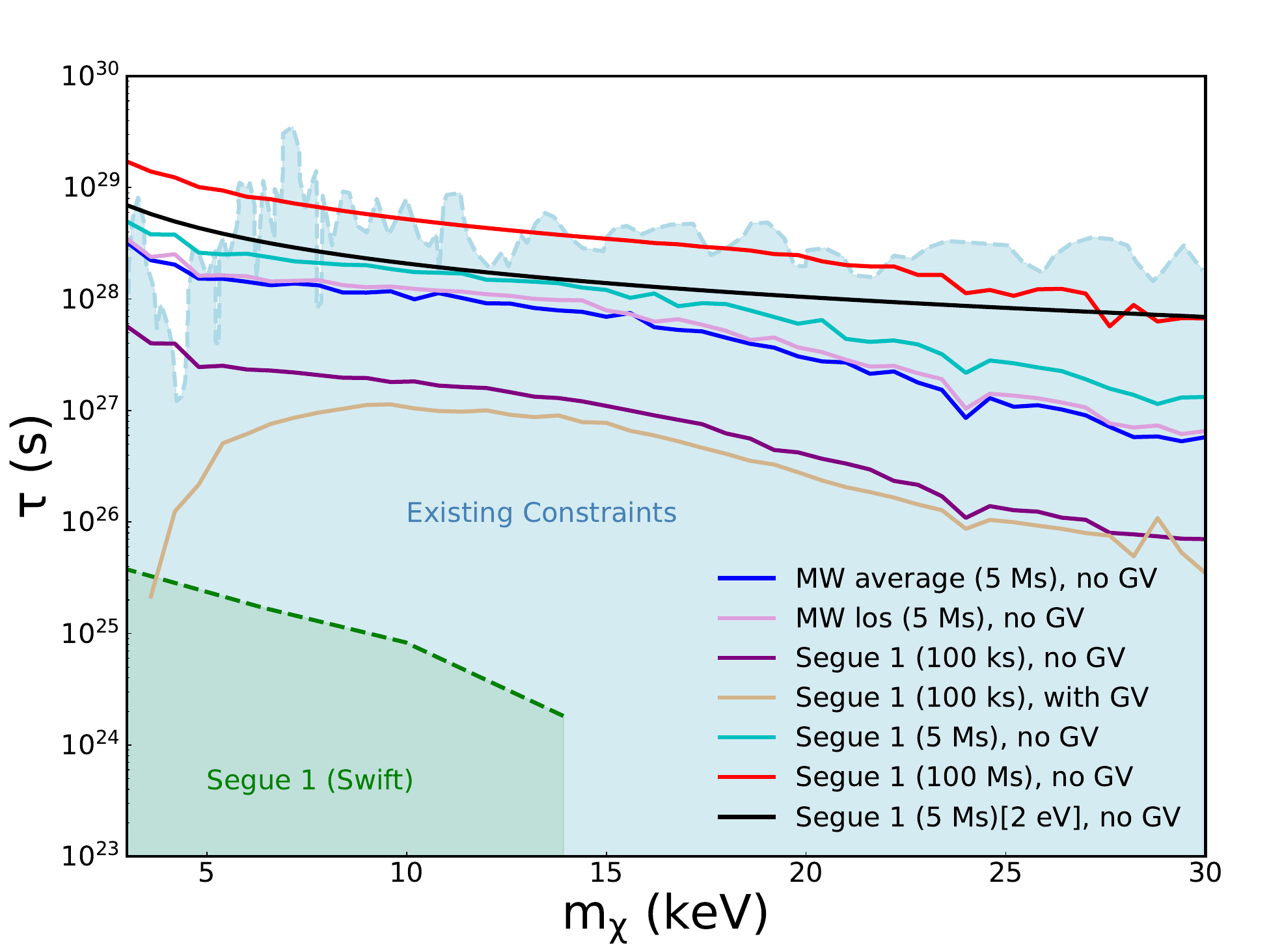}
    \caption{Sensitivity forecast for the lifetime of decaying DM that can be probed by XRISM/Resolve considering various astrophysical target systems, exposure and detector response. 
    Sensitivity curve from Milky Way DM halo for averaged random line-of-sights and assuming gate valve (GV) open (blue), from Milky Way DM halo for the optimized line-of-sight and assuming GV open (pink), from Segue 1 line-of-sight assuming 100 ks observation with GV open (purple), from Segue 1 line-of-sight assuming 100 ks observation with GV closed (brown), from Segue 1 line-of-sight assuming 5 Ms observation with GV open (cyan), from Segue 1 line-of-sight assuming 100 Ms observation with GV closed (red), from Segue 1 line-of-sight assuming 5 Ms observation with a hypothetical detector with 2 eV energy resolution and eight times larger effective area than XRISM/Resolve with GV open (black). 
    Combined existing constraints (blue shaded region) from Ref.~\citep{Horiuchi:2013noa,Dessert:2018qih,Foster:2021ngm,Roach:2019ctw,Roach:2022lgo} as well as constraint (green shaded region) derived from observations of Segue 1 by Swift telescope~\citep{Mirabal2010} are also shown. }
    \label{fig:final sensitivity of decaying time constant} 
\end{figure}

\begin{figure*}
    \centering
    \includegraphics[width=2\columnwidth]{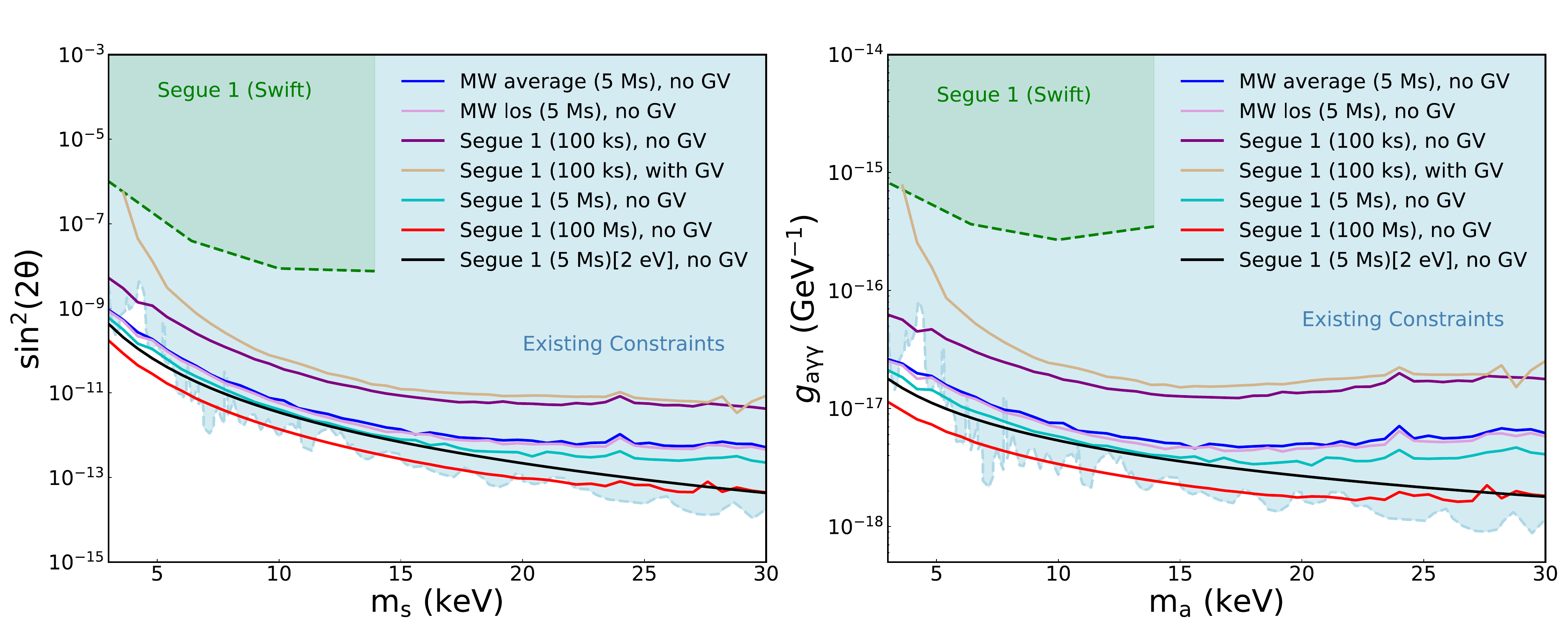}
    \caption{Sensitivity forecast for decaying DM that can be probed by XRISM/Resolve considering various astrophysical target systems, exposure and detector response. [Left] Parameter space for sterile neutrino sterile-active mixing angles. [Right] Parameter space for ALP photon coupling coefficient. Constraints shown are analogous to Fig.~\ref{fig:final sensitivity of decaying time constant}.}
    \label{fig:final sensitivity of active mixing angle and decaying axion coupling constant}
\end{figure*}

\subsection{Sensitivity forecast}
\label{sec:Fit_Results}

The resulting sensitivity projections for decaying DM lifetime are shown in Fig.~\ref{fig:final sensitivity of decaying time constant} considering
various observing conditions including exposure times, gate valve configuration (open or closed) and astrophysical target systems. In Fig.~\ref{fig:final sensitivity of active mixing angle and decaying axion coupling constant} we display sensitivity projections in terms of sterile neutrino sterile-active mixing angle as well as the ALP photon coupling coefficient.

The fact that we used the \texttt{arf} response for flat field may introduce systematic bias in the sensitivity forecast result, given that the DM in Segue 1 is more concentrated in the center region. We estimate the surface brightness deviation caused by using the assumption of the flat field \texttt{arf} response is no more than 40\%. As discussed in Sec.~\ref{section:2.2.3}, the $D$-factors of the Galactic DM halo and Segue 1 DG are comparable. Namely, half of the DM emission along Segue 1 line-of-sight is contributed from the galactic DM halo, for which the flat field response is appropriate. Therefore, overall the systematic bias in the final result should be no more than 20\%.

For the observation target systems we have simulated average of random Galactic DM halo line-of-sights,
optimized Galactic DM halo line-of-sight as well as line-of-sight towards Segue 1. For Segue 1 we compare the sensitivities obtained for 100 ks, 5 Ms, and 100 Ms exposure. Gate valve configuration has a significant impact on the effective area, especially for energies below 2 keV. When the gate valve is closed, sensitivity below 2 keV is suppressed and the effective area decreases by 30\% compared to when the gate valve is open \citep{Tsujimoto2018}. The sensitivity curves display dip features around energy of $\sim 12$~keV, which is equivalent to a DM mass of $\sim 24$~keV. This is mainly corresponding to the dip degradation in the effective area due to the Au absorption.

Intriguingly, our analysis demonstrates that with just 100 ks observation XRISM/Resolve can already improve on the previous Segue 1 DG decaying DM search using Swift data by approximately two orders of magnitude or more in sensitivity to DM coupling and decay time. Observations by XRISM/Resolve with open gate valve enable probing decaying DM, especially with masses below 5 keV, with unprecedented sensitivity. 
Further, we have also analyzed sensitivity of a hypothetical future detector assuming energy resolution of 2 eV. Higher resolution and effective area enable better statistics, which improve the sensitivity primarily at higher energies and DM masses. 

\section{Faint source detection in Segue 1}

Since XRISM/Xtend has a much larger field of view and better effective area than XRISM/Resolve\footnote{Further, XRISM/Xtend is not affected by the status of the gate valve.}, it has a greater potential for discovering faint X-ray point sources. Hereafter our discussion of simulations focuses on XRISM/Xtend.

Segue 1 is a fossil galaxy formed in the early Universe without substantial chemical evolution~\citep{Frebel2014}. Exploring population properties of Segue 1 DG X-ray sources, such as their luminosity function, enables tracing back the formation of sources $\sim1-10$ Gyrs earlier. According to the eROSITA all-sky survey point source catalog~\citep{Merloni2024}, there is an X-ray source located at $\rm R.A.=10h06m46.24s,\ DEC=+16d05m46.02s$, $\sim$4.29 arc minutes away from the Segue 1 center (i.e. $\rm R.A.=10h07m03.2s,\ DEC=+16d04m25s$), with a flux of $3.8\times 10^{-14}\rm\ erg\ cm^{-2}\ s^{-1}$ in the 0.2-2.3 keV range. However, due to the limited exposure of eROSITA observations, it is challenging to identify the energy spectrum of this point source or to establish if it is a member of an X-ray binary system associated with Segue 1. XRISM/Xtend's capabilities enable discovery of the underlying nature of the source with longer exposure time. 

We simulate the image and energy spectrum of the X-ray source in Segue 1
assuming 100 ks observation of XRISM/Xtend and a spectral model using \texttt{TBabs*powerlaw} with $n_{\rm H} = 3.35\times 10^{20}$~cm$^{-2}$ and photon index $\rm PhoIndex = 1.7$. We obtain about 800 net detector counts with a cumulative count rate of $\sim 6.8 \times 10^{-3}\rm\ cts/s$ for this source. Performing a fit with the assumed model yields source parameters within 12\% error of the original assumed values. Fitting to the found simulation spectrum with a thermal bremsstrahlung model, we obtain the best-fit temperature of $11\pm 2$ keV. This result is extremely high compared to the typical energy spectra of low-mass X-ray binary (LMXB) systems with high energy roll-offs at $\sim 2$ keV. Hence, such statistics enables investigating and constraining the true underlying nature of the faint X-ray point source. In particular, LMXB and a background AGN can be clearly distinguished, while high-mass X-ray binary systems (HMXB) with high energy roll-offs typically around $\sim 5 - 10$ keV are marginally distinguishable from AGN. 

\begin{figure*}
    \centering
    \includegraphics[width=2\columnwidth]{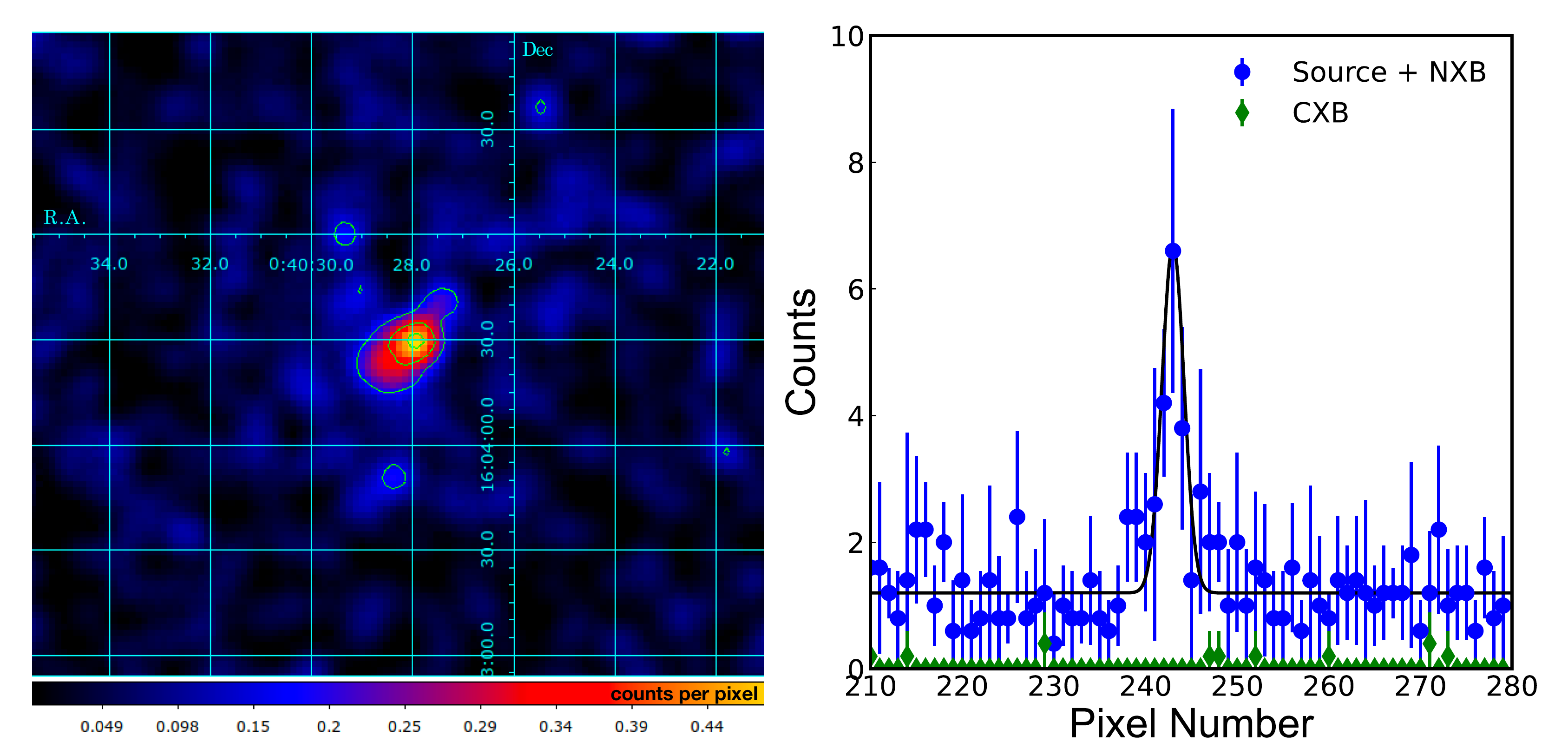}
    \caption{
    [Left] Simulated image of the point source with a flux set at eROSITA all-sky survey detection threshold for XRISM/Xtend 100 ks observation. [Right] Distribution of counts across the detector plane highlighting that the point source counts are well distinguished from the non-X-ray background (flat component across pixels of the blue dots set) and the CXB contribution (green dots).}
    \label{fig:Xtend detection threshold for faint point sources}
\end{figure*}

No X-ray source has been detected within 4 arcmin field from the galaxy center of Segue 1 galaxy at a flux threshold of $1\times 10^{-14}\rm\ erg\ cm^{-2}\ s^{-1}$ for 0.3-10 keV range. If there exists a faint X-ray source with a flux just below the eROSITA survey upper limit at the center of the Segue 1 galaxy, XRISM/Xtend can detect it with $\sim 140$ counts considering 100 ks observation. This is sufficient to distinguish the energy spectrum of the source, thus revealing crucial information for historic star-formation activity in the central galactic region. In Fig.~\ref{fig:Xtend detection threshold for faint point sources} we display the image and a count distribution projected to a side view along the central pixels intersecting the detector plane that covers the presumed X-ray source region. The non-X-ray background and CXB are separately simulated using the calibration input \textit{ah\_sxi\_pch\_nx\_full\_20110530.pi} and the \texttt{Skyback} routine in \texttt{Heasim}. The log$N$-log$S$ relation in the 0.3-10.0 keV range is specified according to in-depth observations \citep{Moretti2003,Moretti2009} for CXB modeling. 
As shown in Fig.~\ref{fig:Xtend detection threshold for faint point sources}, the signal of simulated point source is significantly higher than the non-X-ray background and CXB confusion limit. 

Our analysis demonstrates that XRISM has unique opportunities of discovering new faint X-ray sources in the Segue 1 DG and shed light on underlying historic star-formation activity.

\section{Conclusions}
\label{section:6}

Sensitive X-ray observations of astrophysical systems, such as by the recently launched XRISM satellite mission, can reveal unique insights into the nature of DM as well as the identity of faint X-ray astrophysical sources.
Employing dedicated simulations as well as background and foreground modelling, we comprehensively demonstrated that XRISM can probe decaying DM signatures such as from sterile neutrinos and ALPs, in the few to tens of keV mass-range, with unprecedented sensitivity.

We have identified optimal search strategies for decaying DM, including optimal line-of-sights in case of Galactic DM halo, taking into account
photon foreground emission from the Galactic hot gaseous halo.
Among the DGs, we have identified Segue 1 as a particularly favorable observation target to explore decaying DM signatures considering XRISM/Resolve field of view. Unique recession velocity and internal velocity dispersion of Segue 1 result in a distinctive DM signal along its line-of-sight, which is easily distinguishable from atomic emission lines.
Our forecast analyses of sensitive decaying DM searches by XRISM/Resolve, performed under various assumptions, can be interpreted 
as new probes of sterile-active mixing angle for sterile neutrinos and axion-photon conversion coefficients for ALPs. We find that with just 100 ks observation of Segue 1 XRISM/Resolve 
search can probe parameter space of decaying DM around two orders more sensitively than existing upper limits from Segue 1 established using data collected by the Swift telescope. Further, XRISM/Resolve offers powerful opportunities to probe underexplored DM window around few keV.

Further, we have established XRISM/Xtend's powerful capabilities for discovery of faint X-ray point sources in Segue 1. We found that in-depth observations by XRISM can facilitate identifying the underlying astrophysical nature of these sources, providing valuable insights into the star-forming history of Segue 1.

\acknowledgments

This work was supported by the World Premier International Research Center Initiative (WPI), MEXT,
Japan. V.T. acknowledges support from JSPS KAKENHI grant. No. 23K13109, and K.M. No. 20H05857. This work was performed in part at the Aspen Center for Physics, which is supported
by the National Science Foundation grant PHY-2210452.
 
\bibliographystyle{aasjournal}
\bibliography{refs}

\end{document}